\documentclass{epl}
\usepackage[latin2]{inputenc}
\usepackage{amsfonts}
\usepackage{amsmath}
\usepackage{amssymb}
\usepackage{graphics,epsf}
\newcommand{\f}{\widetilde{\Phi}(\E,M)}
\newcommand{\E}{\mathcal{E}}
\newcommand{\I}{\text{i}}
\title{Fermi-Dirac statistics and the number theory}
\author{Anna Kubasiak\inst{1} \and Jaros\l{}aw K. Korbicz\inst{2} \and Jakub
Zakrzewski\inst{1}, \and Maciej Lewenstein\inst{2,3,*}}
\institute{
  \inst{1} Instytut Fizyki im. M. Smoluchowskiego, Uniwersytet Jagiello\'nski, PL-30059 Krak\'ow, Poland\\
  \inst{2} Institut f\"ur Theoretische Physik, Universit\"at Hannover, D-30167 Hannover, Germany\\
  \inst{3} ICFO-Institut de Ci\`encies Fot\`oniques, E-08034 Barcelona, Spain
}
\pacs{03.75.Ss}{Degenerate Fermi gases}
\pacs{05.30.Fk}{Fermion systems and electron gas}
\pacs{02.30.Mv}{Approximations and expansions}

\begin{document}

\maketitle

\begin{abstract}
We relate the Fermi-Dirac statistics of an ideal Fermi gas in a harmonic trap to partitions of 
given integers into distinct parts, studied in number theory. Using methods of quantum statistical 
physics we derive analytic expressions for
cumulants of the probability distribution of the number of different partitions.   
\end{abstract}

\section{Introduction}
Standard textbooks use grand canonical ensemble to describe {\it ideal} quantum  gases 
\cite{Huang}. It has 
been pointed out by Grossmann and Holthaus \cite{Grossmann}, and Wilkens et al. \cite{martin1}
 that such an approach leads to  pathological results for 
fluctuations of particle numbers in Bose-Einstein condensates (BEC). This has stimulated intensive studies 
of BEC fluctations using canonical and microcanical ensembles  for ideal \cite{studies,Maxwell} and weakly 
interacting \cite{studiesweak} gases. Grossmann and Holthaus \cite{Grossmann} related the problem for an 
ideal Bose gas in an harmonic trap  to the number theoretical studies of 
partitions of an integer $\E$ into $M$ integers \cite{Andrews}, and to the famous Hardy-Ramanujan formula 
\cite{Hardy}. In the series of beautiful papers, they have been able to apply methods of 
quantum statistics to derive non-trivial number theoretical results (see \cite{Holthaus} and references therein). In this Letter we apply the 
approach of Ref. \cite{Holthaus} to Fermi gases, and relate the problem of an ideal Fermi gas 
in an harmonic trap to studies of partitions of $\E$ into $M$ {\it distinct} integers. 
We show that the probability distribution of the number of different partitions 
is asymptotically Gaussian and we calculate analytically its cumulants. Our results are complementary
to those of Tran \cite{tran}, who has calculate particle number fluctuations in the ground state (Fermi sea)
for the fixed energy and number of particles.

\section{Ideal Fermi gas}
We consider a microcanonical ensemble of $N$ thermally isolated, spinless, non-interacting fermions, 
trapped in a harmonic potential with frequency $\omega_0$. The discrete one-particle states have 
energies $\epsilon_\nu=\hbar \omega_0(\nu+1/2)$, $\nu=0,1,\dots$. A microstate of the system is described 
by the sequence of the occupation numbers $\{n_\nu\}$, where $n_\nu=0,1$ and $\sum_{\nu=0}^\infty n_\nu=N$. 
The total energy of a microstate $\{n_\nu\}$ can be written as:
\begin{equation}\label{energia}
E(\{n_\nu\})=N\epsilon_0+\hbar \omega_0\, \E \, ,
\end{equation}
where integer:
\begin{equation}
\E=\sum_{\nu=1}^\infty n_\nu \, \nu
\end{equation}
determines the contribution to $E$ from one-particle excited states. Note, that $\E$ is a sum of $M$ distinct integers, where $M=N-1$ if $n_0=1$, or $M=N$ otherwise. Hence, in order to construct the microcanonical partition function $\Gamma(E,N)$, one has to calculate the number of partitions of the given integer $\E$ into $M$ distinct parts. This, in turn, is a standard problem in combinatorics \cite{combi}. If we call the number of distinct partitions $\widetilde{\Phi}(\E,M)$, then $\Gamma(E,N)=\f$ and the physical problem of finding $\Gamma(E,N)$ is mapped to a combinatorial one.

Introduce the total number of distinct decompositions:
\begin{equation}
\widetilde{\Omega}(\E)=\sum_{M=1}^{M_{max}}\f,
\end{equation}
and consider the probability (relative frequency):
\begin{equation}
p_{mc}(\E,M)=\frac{\f}{\widetilde{\Omega}(\E)}\, , 
\label{pmc}
\end{equation}
that exactly $M$ distinct terms occur in the decomposition of $\E$. This is the central object of our study. 
In the case of bosons $p_{mc}(\E,M)$ possesses clear physical interpretation, as it was explained in Ref. \cite{Holthaus}. This is the probability of finding $M$ excited particles, when the total excitation energy is $\hbar\omega_0\,\E$. For fermions, however, the ground state of the whole system corresponds to the filled Fermi sea. Since the number of partitions $M$ differs at most by $1$ from the number of particles $N$, we may regard (\ref{pmc}) as a probability density in the fictitious Maxwell demon ensemble \cite{Maxwell} with vanishing chemical potential. The distribution (\ref{pmc}) describes at the same time an  interesting mathematical problem, which, up to our knowledge, has never been treated using physical methods. 

\begin{figure}
\includegraphics[width=0.75\linewidth]{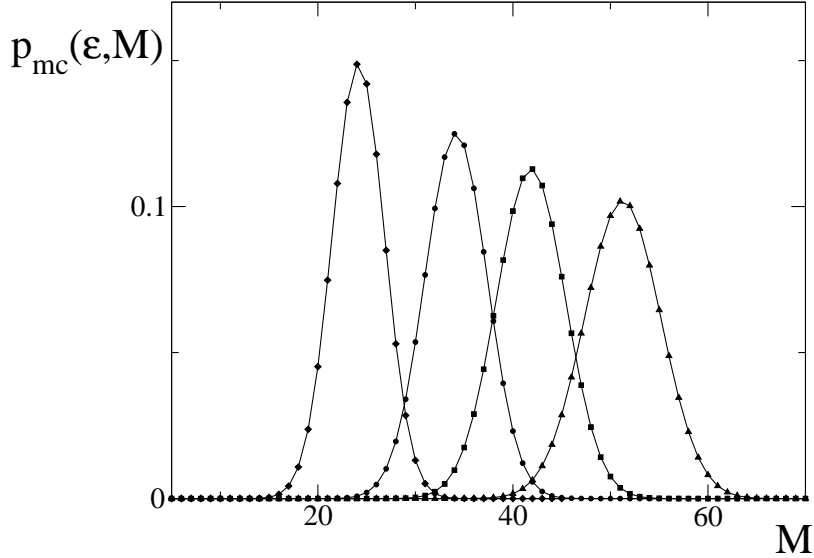}
\caption{The distribution $p_{mc}(\E,M)$ for $\E=1000,2000,3000$, and $4500$.}
\label{distr}
\end{figure}

Using standard results from combinatorics \cite{combi}, the exact expression for $\f$ can be written down: 
\begin{equation}
\f = P\bigg(\E-{M\choose 2},M\bigg)\, , 
\label{rek}
\end{equation}
in terms of a recursively defined function $P$:
\begin{equation}\label{P}
P(n,k)=\left\{\begin{array}{l}  0 \ \ \ \mbox{for} \ \ k=0 \ \ \mbox{or} \ \ k> n\\
                                P(n-1,k-1)+P(n-k,k) \\
                                1 \ \ \ \mbox{for}\ \  k=n \end{array}\right.\,.
\end{equation}
The above recursion  enables straightforward,  although tedious, numerical studies of the distribution (\ref{pmc}) (see  Fig. \ref{distr}).
 
\section{Canonical distribution}
To proceed further with the analytical study of $p_{mc}(\E,M)$ we introduce the corresponding canonical description.
 Instead of fixing the energy $E$ of the each member of the ensemble, we consider an ensemble of systems in a thermal equilibrium with a heat bath of temperature $T$. The canonical partition function $Z(\beta,N)$, $\beta=1/kT$, generating the statistical description in this case, can be written using (\ref{energia}) as: 
\begin{equation}
Z(\beta,N)=\sum_{E}\text{e}^{-\beta E}\Gamma(E,N)
=\text{e}^{-\beta N \epsilon_0}\sum_{\E=0}^\infty \text{e}^{-\hbar\omega_0 \beta \E}\, \f\,,\label{Zcanon}
\end{equation}
where $\E=0$ term vanishes.
Introducing $b=\hbar\omega_0 \beta$ we can define a combinatorial partition function:
\begin{equation}\label{Zc}
Z_{c}(b,M)=\sum_{\E=1}^\infty \text{e}^{- b \E}\, \f\, ,
\end{equation}
which can be viewed as the generating function of the microcanonical quantities $\f$. The canonical probability, corresponding to (\ref{pmc}), is then given by:
\begin{equation}
p_{cn}(b,M)=\frac{Z_{c}(b,M)}{\sum_{\E=1}^\infty \text{e}^{- b \E}\,\widetilde{\Omega}(\E)}\, , 
\label{pcn}
\end{equation}
A convenient and physically relevant way of an analytical description of the distribution (\ref{pcn}) is provided by the cumulants. For a probability distribution $\{p(M),M\ge 0\}$ of an integer-valued random variable, $k$th cumulant $\kappa^{(k)}$ is defined as \cite{Abramowitz}:
\begin{equation}\label{kumul}
\kappa^{(k)}=\bigg(z \frac{d}{dz}\bigg)^k\text{ln}\,\hat{p}(z)\bigg| _{z=1}\, ,
\end{equation}
where $\hat{p}(z)=\sum_M \text{e}^{-Mz}\, p(M)$ is the (real) characteristic function of the distribution.  The relation of the first few culumants to the central moments $\mu^{(k)}$ is the following:
\begin{equation}
\kappa^{(1)}=\mu^{(1)}=\bar{M} \quad\quad\quad \kappa^{(2)}= \mu^{(2)} \quad\quad\quad \kappa^{(3)}= \mu^{(3)} \quad\quad\quad \kappa^{(4)}= \mu^{(4)}-3(\mu^{(2)})^2\ ,
\end{equation}
where $\bar{M}=\sum_M M\, p(M)$ is the mean value. Denote as 
$
\Xi_c (b,z)=\sum_{M=0}^\infty z^M Z_{c}(b,M)
=\prod_{\nu=0}^\infty (1+z\text{e}^{-b \nu})
$
the grand partition function, corresponding to the canonical partition function (\ref{Zc}) with  $z$ being an analog of fugacity. 
We are interested in excited part  so we take not $\Xi_c(b,z)$ but rather $\Xi_{ex}(b,z)$.
\begin{equation}
\Xi_{ex}(b,z)=\prod_{\nu=1}^\infty \big(1+z\text{e}^{-b \nu}\big) .\label{Xiex}
\end{equation}
We substitute (\ref{pcn}) in to the definition (\ref{kumul}) and obtain:
\begin{equation}\label{kcn}
\kappa^{(k)}_{cn}(b)=\bigg(z \frac{\partial}{\partial z}\bigg)^k\text{ln}\,\Xi_{ex} (b,z)\bigg| _{z=1}\,.
\end{equation}

Now, we can apply the procedure of Ref. \cite{Holthaus} in order to obtain a compact expression for the cumulants $\kappa^{(k)}_{cn}(b)$ as well as their asymptotic behavior for $b\ll 1$. First, we note that:
\begin{equation}
\ln{\Xi_{ex} (b,z)}=\sum_{\nu=1}^{\infty}\ln\big(1+z\text{e}^{-b \nu}\big)\,.
\end{equation}
and we can always find such a neighborhood of $z=1$ that $z\text{e}^{-b}<1$. Since, according to (\ref{kcn}), we are only interested in the behavior of $\Xi_{ex}$ in the vicinity of $z=1$, the logarithms in the second term can be Taylor expanded; using then the Mellin transformation of the Euler gamma function $\Gamma(t)$ \cite{Mellin}, we 
obtain the following integral representation of $\ln \Xi_{ex}$ :
\begin{equation}\label{Xiex4}
\ln\Xi_{ex}(b,z)= \frac{1}{2\pi \I}\int_{\tau-\I\infty}^{\tau+\I\infty}dt\ b^{-t} \zeta(t)\Gamma(t)f_{t+1}(z)\,,
\end{equation}
where $\zeta(t)$ is the Riemann zeta function \cite{Abramowitz}, and 
\begin{equation}\label{f}
f_{\alpha}(z)=\sum_{n=0}^\infty \frac{(-1)^n z^n}{n^\alpha} .
\end{equation}
is the fermionic function \cite{Huang}. The parameter $\tau >0$ was chosen in such a way, that the integration line in (\ref{Xiex4}) is to the right from the last pole of the integrand. 

Finally, we substitute Eq. (\ref{Xiex4}) into (\ref{kcn}) and with the help of the following properties of the fermionic function $f_\alpha$:
$z\,df_{\alpha}(z)/dz=f_{\alpha-1}(z)$, 
$f_{\alpha}(1)=(1-2^{1-\alpha})\zeta(\alpha)$, 
we find that:
\begin{equation}\label{kcn2}
\kappa_{cn}^{(k)}(b)= \kappa_{cn,boson}^{(k)}(b)-2^k\kappa_{cn,boson}^{(k)}(2b)\,,
\end{equation}
where 
\begin{equation}\label{kboson}
\kappa_{cn,boson}^{(k)}(b)=\frac{1}{2\pi \I}\int_{\tau-\I\infty}^{\tau+\I\infty}dt\ b^{-t} \Gamma(t)\zeta(t) \zeta (t+1-k)
\end{equation}
are the canonical cumulants of the bosonic counterpart of the distribution (\ref{pcn}).

Equation (\ref{kcn2}) is of central importance for this letter. It links the fermionic
cumulants to that corresponding for bosons. The latter has been studied in detail in Ref. \cite{Holthaus}. In particular \cite{Holthaus} gives  the asymptotic expressions of the first few bosonic cumulants for $b \ll 1$. Using those together with (\ref{kcn2}-\ref{kboson})
 allow us to find the asymptotic expressions for the cumulants of the fermionic canonical distribution (\ref{pcn}) :
$
\kappa_{cn}^{(0)}(b)= {\pi^2}/{12 b}+\ln 2/2 +{b}/{24}+\mathcal{O}(b^{15.5})$, 
$\kappa_{cn}^{(1)}(b)= {\ln 2}/{b}-{1}/{4}+{b}/{48}+\mathcal{O}(b^{2.5})$,
$\kappa_{cn}^{(2)}(b)= {1}/{2b}-{1}/{8}+\mathcal{O}(b^{14.5})$,
$\kappa_{cn}^{(3)}(b)= {1}/{4b}-{b}/{96}+\mathcal{O}(b^{2.5})$,
$\kappa_{cn}^{(4)}(b)= -{1}/{16}+\mathcal{O}(b^{11.5})$, etc.
The dependence on $b$ in the fermionic case is of lower order than the correspponding one in the bosonic case. For example $\kappa_{cn}^{(4)}$ is constant up to the terms of the order $b^{11.5}$.

\section{Microcanonical distribution}

We come back to the study of the distribution (\ref{pmc}). Our aim now is to find the cumulants  $\kappa_{mc}^{(k)}(\E)$ of this distribution. Applying the definition (\ref{kumul}) to (\ref{pmc}) we find that:  
\begin{equation}
\kappa_{mc}^{(k)}(\E)=\left(z\frac{\partial}{\partial z}\right)^k\ln\Upsilon(\E,z)\bigg|_{z=1}\,,
\label{kmc}
\end{equation} 
where:
\begin{equation}\label{Y}
\ln\Upsilon(\E,z)=\sum_{M=0}^\infty z^M \f\,
\end{equation}
is another generating function of $\f$, complementary  to $Z_c(b,M)$.

In order  to calculate $\Upsilon(\E,z)$, we first note, that:
\begin{equation}
\Xi_{ex} (b,z)=\sum_{M=0}^\infty z^M \sum_{\E=1}^\infty \text{e}^{- b\E}\, \f
=\sum_{\E=1}^\infty \text{e}^{- b\E}\Upsilon(\E,z)\label{Y2}\,. 
\end{equation}
Introducing a new variable $x=\text{e}^{-b}$ and witting $\Xi_{ex}(-\ln x,z)=\widetilde{\Xi}_{ex}(x,z)$ we obtain:
\begin{equation}
\widetilde{\Xi}_{ex}(x,z)=\sum_{\E=1}^{\infty}x^\E\Upsilon(\E,z)\,,
\end{equation}
that is, $\Upsilon(\E,z)$ are the coefficients in the power series expansion of $\widetilde{\Xi}_{ex}(x,z)$ w.r.t. $x$. We treat $x$ as a complex variable and use an integral identity: 
\begin{equation}\label{Y3}
\Upsilon(\E,z)=\frac{1}{2\pi \I}\oint dx \frac{\widetilde{\Xi}_{ex}(x,z)}{x^{\E+1}},
\end{equation}
where the integration contour is any closed loop surrounding the origin, oriented anti-clockwise. The integral in (\ref{Y3}) can be approximately evaluated using the saddle point method similar to that used in Refs. \cite{Holthaus, Huang}. The implicit equation for the saddle point $x_0(z)$ is:  
\begin{equation}\label{sp2}
\E+1 = -\frac{\partial}{\partial b}\ln \Xi_{ex}(b_0(z),z)\ , 
\end{equation}
with $b_0(z)=-\ln x_0(z)$. From Eq. (\ref{sp2}) we obtain in the Gaussian approximation:
\begin{equation}
\ln\Upsilon(\E,z)\approx\ln \Xi_{ex}(b_0(z),z)+\E b_0(z)
-\frac{1}{2}\ln 2\pi-\frac{1}{2}\ln \bigg[\left( -\frac{\partial}{\partial b}\right)^2 \!\!\ln \Xi_{ex}(b_0(z),z)\bigg]\,.
\label{lnY}
\end{equation}
The above solution for $\Upsilon(\E,z)$ is of course similar to the one obtained in a bosonic case in Ref. \cite{Holthaus}, the difference being just in the form of the grand canonical partition function $\Xi_{ex}$.

\begin{figure}
\includegraphics[width=0.7\linewidth,clip=1cm]{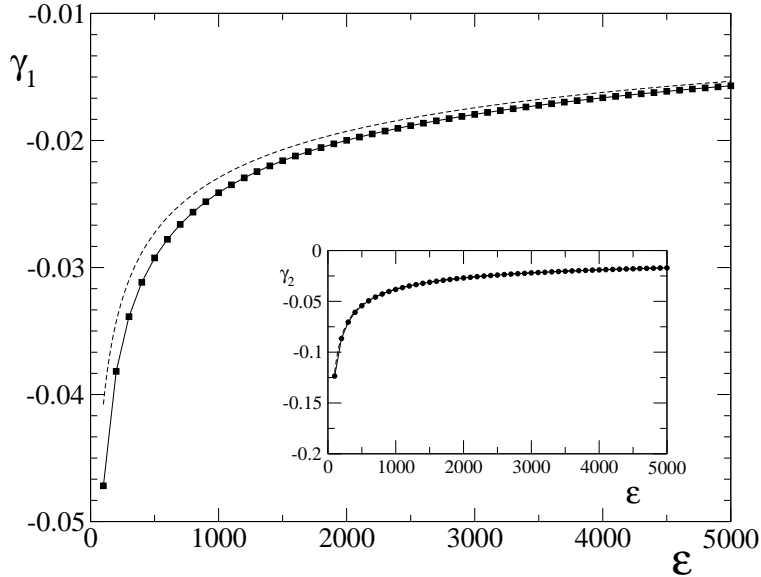}
\caption{The skewness $\gamma_1$ obtained from numerically generated cumulants
as compared with the asymptotic analytic prediction (dashed line).   The insert shows the comparison for the excess $\gamma_2$. Observe that the analytical asymptotics (dashed line) practically coincides with the numerical data.}
\label{excess}
\end{figure}

The microcanonical cumulants are then calculated from (\ref{lnY}) according to the equation (\ref{kmc}):
\begin{equation}
\kappa_{mc}^{(k)}(\E)=\left(z\frac{\partial}{\partial z}+ z \frac{d b_0(z)}{dz}\frac{\partial}{\partial b_0}\right)^k\ln\Upsilon(\E,z)\bigg|_{z=1}\, ,\label{kmc2}
\end{equation}
where from Eq. (\ref{sp2}) 
$
{d b_0(z)}/{dz}\bigg|_{z=1}= -{\partial_{b}\kappa_{cn}^{(1)}(b_0(1))}/{\partial_{b}^2\kappa_{cn}^{(0)}(b_0(1))} $,
where in the last expression  we used the definitions of the canonical cumulants (\ref{kcn}). 
Generally, from (\ref{kcn}),(\ref{lnY}) and  (\ref{kcn})  it follows that the microcanonical 
cumulants $\kappa_{mc}^{(k)}(\E)$ can be expressed in terms of the canonical 
cumulants $\kappa_{cn}^{(k)}(b)$ and their derivatives w.r.t. the temperature parameter $b$. 
For example for the first cumulant $\kappa_{mc}^{(1)}$ one obtains:
\begin{equation}
\kappa_{mc}^{(1)}(\E)=\kappa_{cn}^{(1)}(b_0(1))-\frac{1}{2}\frac{\partial_{b}^2\kappa_{cn}^{(1)}(b_0(1))}{\partial_{b}^2\kappa_{cn}^{(0)}(b_0(1))}+\frac{\partial_{b}\kappa_{cn}^{(1)}(b_0(1))}{\partial_{b}^2\kappa_{cn}^{(0)}(b_0(1))}\left(1+\frac{1}{2}\frac{\partial_{b}^3\kappa_{cn}^{(0)}(b_0(1))}{\partial_{b}^2\kappa_{cn}^{(0)}(b_0(1))}\right)\,.
\end{equation}
For $b\ll 1$ we can use calculate the first few microcanonical cumulants explicitly: 
{\setlength \arraycolsep{2pt}
\begin{eqnarray}
\kappa_{mc}^{(1)}(\E)&=&\frac{2\sqrt{3}\ln 2}{\pi}\sqrt{\E}-\frac{1}{4}+\frac{3\ln 2}{\pi^2} +\mathcal{O}(\E^{-\frac{1}{2}}) \label{K1}\\ 
\kappa_{mc}^{(2)}(\E)&=& -\frac{\sqrt{3}}{\pi^3}\big[-\pi^2+12(\ln2)^2\big]\sqrt{\E}-\frac{36(\ln2)^2-\frac{2}{3}\pi^2}{\pi^4}-\frac{1}{8}+\mathcal{O}(\E^{-\frac{1}{2}})\label{K2}\\ 
\kappa_{mc}^{(3)}(\E)&=& \frac{\sqrt{3}}{\pi^5}\big[-36\pi^2\ln2+\pi^4+432(\ln2)^3\big]\sqrt{\E}\nonumber \\
&&+\frac{\frac{3}{4}\pi^4-54\pi^2+864(\ln2)^3}{\pi^6}\ +\mathcal{O}(\E^{-\frac{1}{2}})\label{K3}\\ 
\kappa_{mc}^{(4)}(\E)&=&-\frac{3\sqrt{3}}{\pi^7} \big[ 4\pi^4\ln2+2160(\ln2)^4-216\pi^2(\ln2)^2+3\pi^4\big]\sqrt{\E}\nonumber \\
&&+\frac{1}{\pi^8}\Big[\frac{1}{16}\pi^8+31104(\ln2)^4+27\pi^4-2592\pi^2(\ln2)^2+36\pi^4\ln2\Big]+\mathcal{O}(\E^{-\frac{1}{2}})\label{K4}\,.
\end{eqnarray}}
Using (\ref{K1}-\ref{K4}) we obtain analytic expressions for the  skewness and excess 
\cite{Abramowitz} of the distribution (\ref{pmc})  
$
\gamma_1(\E)= {\kappa_{mc}^{(3)}(\E)}/{(\kappa_{mc}^{(2)}(\E))^{\frac{3}{2}}}$,
$\gamma_2(\E)= {\kappa_{mc}^{(4)}(\E)}/{(\kappa_{mc}^{(2)}(\E))^2}\label{g2}$;
these parameters measure the deviation of the distribution (\ref{pmc}) from the Gaussian one. We obtain:
{\setlength \arraycolsep{2pt}\begin{eqnarray}
\gamma_1(\E)&=&-\frac{0.12894}{^4\sqrt{\E}}+\mathcal{O}(\E^{-\frac{3}{4}})\nonumber
\\
\gamma_2(\E)&=&-\frac{1.2001}{\sqrt{\E}}+\mathcal{O}(\E^{-1})\,,
\label{g2a}
\end{eqnarray}}
and hence in the limit $\E\to \infty$ both $\gamma_1$ and $\gamma_2$ vanish, implying that for 
large integers $\E$ the distribution (\ref{pmc}) approaches Gaussian. This is a different situation 
from the bosonic case, where, as shown in Ref. \cite{Holthaus}, both skewness and excess 
 are non-zero in the limit of large integers, so that bosonic microcanonical distribution 
does not approach Gaussian. In Fig.\ref{excess} we compare the numerically obtained dependence of the skewness
and excess with the analytical predictions of Eq. (\ref{g2a}).

Summarizing, we have studied relations between the Fermi-Dirac statistics for an ideal, 
harmonically trapped Fermi gas,  and the theory of partitions of integers into distinct parts. 
Using methods of quantum statistical physics,  we have described analytically properties of the 
probability distribution of the number of different partitions. We anticipate that the results 
concerning microcanonical ensemble for fermions can be used to characterize degeneracies of 
fermionic many body states, that  play an essential role in the process of laser cooling of
 small samples of atoms in microtraps  in the so called Lamb-Dicke limit \cite{degenerat}. 
  
We acknowledge discussions with M. Holzmann, and support from the Deutsche 
Forschungsgemeinschaft (SFB 407, 432 POL),  EU IP "SCALA" and the Socrates Programme. This work was
supported (A.K.) by Polish government scientific funds (2005-2008) as a research grant. Support of
Polish Scientific funds PBZ-MIN-008/P03/2003 is also acknowledged (J.Z.).

\end{document}